\documentclass[fleqn,usenatbib]{mnras}

\usepackage{newtxtext,newtxmath}

\usepackage[T1]{fontenc}

\DeclareRobustCommand{\VAN}[3]{#2}
\let\VANthebibliography\thebibliography
\def\thebibliography{\DeclareRobustCommand{\VAN}[3]{##3}\VANthebibliography}

\usepackage{graphicx}	
\usepackage{amsmath}	
\graphicspath{{Images/}}

\newcommand{\jup}{\,{\rm M}_{\rm J}} 
\newcommand{\msol}{\,{\rm M}_{\odot}} 

\title[The dynamical stability of wide-orbit Jupiters]{
On the survivability of a population of gas giant planets on wide orbits}

\author[E. J. Carter, D. Stamatellos]{
Ethan J. Carter,$^{1}$\thanks{E-mail: ecarter6@uclan.ac.uk}
Dimitris Stamatellos,$^{1}$\thanks{E-mail: dstamatellos@uclan.ac.uk}
\\
$^{1}$Jeremiah Horrocks Institute for Mathematics, Physics \& Astronomy, University of Central Lancashire, Preston PR1 2HE, UK\\
}

\date{Accepted 2023 July 25. Received 2023 July 07; in original form 2023 May 17}

\pubyear{2023}

\begin{document}
\label{firstpage}
\pagerange{\pageref{firstpage}--\pageref{lastpage}}
\maketitle

\begin{abstract}
 The existence of giant planets on wide orbits ($\stackrel{>}{_\sim}100\rm~AU$) challenge planet formation theories; the core accretion scenario has difficulty in forming them, whereas the disc instability model forms an overabundance of them that is not seen observations. We perform $N$-body simulations investigating the effect of close stellar encounters ($\leq 1200$~AU) on systems hosting wide-orbit giant planets and the extent at which such interactions may disrupt the initial wide-orbit planet population. We find that the effect of an interaction on the orbit of a planet is stronger for high-mass, low-velocity perturbers, as expected. We find that due to just a single encounter there is a $\sim 17\%$ chance that the wide-orbit giant planet is liberated in the field, a $\sim 10$\% chance it is scattered significantly outwards, and a $\sim 6$\% chance it is significantly scattered inwards. Moreover, there is a $\sim 21\%$ chance that its eccentricity is excited to e>0.1, making it more prone to disruption in subsequent encounters. The results strongly suggest that the effect of even a single stellar encounter is significant in disrupting the primordial wide-orbit giant planet population; in reality the effect will be even more prominent, as in a young star-forming region more such interactions are expected to occur. We conclude that the low occurrence rate of wide-orbit planets revealed by observational surveys does not exclude the possibility that such planetary systems are initially abundant, and therefore the disc-instability model may be a plausible scenario for their formation.
 
\end{abstract}




\section{Introduction}
The properties of exoplanetary systems are diverse and quite different with those of our own solar system. We must look to the architecture of other planetary systems to better understand the mechanisms by which our own solar system formed. 
Whilst the majority of known exoplanets are observed at small separations from their host, several giant exoplanets have been confirmed on wide orbits \citep{Marois:2008a, Marois:2010a, Bailey:2013, Chauvin:2017, Janson:2021, Gaidos:2021, Zhang:2021,Fontaive:2020, Bohn:2020}. Direct imaging surveys show that only a small percentage of stars host gas giant planet on wide orbits; up to a maximum of $\sim 10\% $ of stars, with a small dependence on the stellar host mass \citep{Brandt:2014a, Bowler:2015b, Lannier:2016a, Reggiani:2016a,Galicher2016, Bowler:2016a, Vigan:2017a, Baron2018a, Stone:2018a,Wagner:2019w, Nielsen:2019a} \citep[see review by][]{Bowler:2018a}. The most recent survey from \cite{Vigan:2021w} has found that the frequencies of stars with at least one massive substellar companion ($1-75~{\rm M_J}$) at distance from 5 to 300 AU is $23.0^{+13.5}_{-9.7}$, $5.8^{+4.7}_{-2.8}$,  and $12.6^{+12.9}_{-7,1}$, for BA, FGK, and M stars, respectively. The existence of gas giants on such wide orbits challenges current planet formation theories.

There are two primary theories for planet formation: (i) core accretion and (ii) disc fragmentation due to gravitational instabilities. Core accretion proposes that a core of rock or ice forms by pebble or planetesimal accretion fast enough that it can accrue gas rapidly forming a large, heavy element-rich, gaseous planet after reaching a critical core mass \citep{Pollack:1996,Lambrechts:2012k}. The core accretion mechanism is thought to form giant planets optimally in the $5-10~\rm AU$ region around a $1\rm M_{\odot}$ star \citep{Helled:2014}, and struggles to explain the existence of giant planets at distances greater than $~20~\rm AU$. The model requires a few million years to form gas giants, a timescale likely longer than the lifetime of protoplanetary discs \citep{Hernandez:2008}. As a result, the formation of gas giants on wide-orbits, especially those around M dwarfs \citep[e.g.][]{Morales:2019a} poses an issue for the core accretion formation theory. 

The disc instability theory is capable of explaining the formation of giant planets on wide orbits \citep{Stamatellos:2009a,Helled:2014,Mercer:2020a}. This model describes the concept that massive protoplanetary discs may fragment to form planets due to gravitational instability as a result of their self-gravity \citep{Kuiper:1951,Cameron:1978, Boss:1997a}. There are two prerequisites for fragments to form via disc instability(i) the disc must be massive enough for gravity to dominate over thermal and centrifugal support \citep{Toomre:1964}, and (ii) the disc must cool fast enough \citep[on a dynamical timescale;][]{Gammie:2001,Rice:2005}. In this scenario, fragments form in the outer disc region where the two criteria are satisfied at the same time $(\geq 50-100\rm AU)$ \citep{Stamatellos:2007, Stamatellos:2009a, Stamatellos:2009d, Boley:2009a}. Disc fragmentation is expected to happen when the disc is young and therefore relatively massive compared to the host star with a disc to mass ratio \citep[e.g][]{Cadman:2020, Mercer:2020a}, { with the exact outcome depending on the specific disc properties, such as metallicity temperature and size \citep{Meru:2010a}.} However, observations show that the most common outcome of disc instabilities, i.e. massive gas giants on wide orbits are not very common \citep[e.g.][]{Vigan:2021w,Rice:2022b}.
{ Based on this, \cite{Rice:2015a} argue that disc fragmentation rarely forms planetary-mass objects, whereas 
\cite{Nayakshin:2017b} argues that an initial abundant population of such planets effectively disappears due to inward migration, mass growth or tidal disruption.} 

Planetary systems typically form in clusters or open associations as opposed to in isolation and are susceptible to the influence of surrounding stars \citep{Lada:2003}. Planets in a cluster environment may have their orbits altered by dynamical encounters such as close stellar flybys {\citep{Thies:2011a, Parker:2012, Perets:2012a, Hao:2013a, Zheng:2015g,Flammini-Dotti:2019, Jimenez:2020,Parker:2020}}. The orbits of wide-orbit giant planets may be hardened, i.e. planets are scattered inwards, contributing to the observed hot Jupiter population {\citep{Wang:2022, Li:2023}}, or strong gravitational perturbations may soften the orbit leading to an eventual ejection of the planet from its host star, contributing to the population of free-floating planets \{\citep[e.g. ][]{Miret-Roig:2022, Hurley:2002}. {A planet that has been ejected from its host star may be captured by a new star \citep[e.g. ][]{Perets:2012a, Wang:2015, Cai:2019, Fujii:2019}, or directly exchanged between stars as they pass each other \citep{Daffern-Powell:2022, Mustill:2016, Wang:2020, Wang:2020b}.}

Dynamical interactions between stars and their planetary companions within young stellar clusters have been a focus of investigation as a prominent mechanism for the shaping of planetary systems as we observe them today. \cite{Pfalzner:2018} proposed that dynamical interactions with a passing star may have shaped the young solar system, and suggested that a close stellar flyby could recreate the prominent characteristics of our solar system as observed today. Recently, \cite{Miret-Roig:2022} discovered of $70-170$ free-floating planets in the region encompassing Upper Scorpius and Ophiuchus, a population higher than expected from the turbulent fragmentation theory. They suggest that ejections due to dynamical instabilities in planetary systems hosting giant planets must be frequent in the first $10~\rm Myr$ of the system's life. 


In this paper, we perform $N$-body simulations of planetary systems perturbed by passing stars (as expected in a cluster environment) to determine the significance of close stellar encounters in shaping the observed population of wide orbit planets and in contributing to the population of free-floating planets. More specifically we examine the dynamics of a wide-orbit Jupiter-mass planet placed on an initially circular orbit around a host star as the planet-star system is perturbed by a passing star. Our goal is to explore how a single interaction may alter the architecture of the system. The main question that we will try to answer is whether the significant initial wide orbit planet population that is predicted by the disc fragmentation theory is able to survive long term within a cluster environment.

In section 2 we describe the details of the computational method that we use, and in Section 3 the initial setup of the planetary system and the free parameters of our study. In Section 4 we present our results regarding the dynamical stability of wide orbit Jupiters and, and in Section 5 we discuss how these depend on the host mass, and the perturber mass, velocity, impact parameter and direction of approach. Finally, in Section 6 we place our results within the context of planet formation theories.

\section{Computational Methods}

We simulate the dynamical evolution of a planetary system with a giant planet on a wide orbit as this is perturbed by a passing star using an $N$-body code \citep{Hubber:2005, Hubber:2011c}, which utilises a fourth-order Hermite integration scheme. 

In a Hermite time-step scheme, body $i$ has a position $x_{i}$ and a velocity $v_{i}$ at time $t_{i}$. The $N$-body code adopts a global time-step $\Delta t_{i}$, so that
\begin{equation}
\Delta t_{i} = \gamma \sqrt{\frac{| \textbf{a}_{i} | | \ddot{\textbf{a}_{i}} | + | \dot{\textbf{a}^{2}_{i}} |}{{| \dot{\textbf{a}_{i}} || \dddot{\textbf{a}_{i}}| + | \ddot{\textbf{a}^{2}_{i}} |}}}
\end{equation}
where $\dot{\textbf{a}_i},\ddot{\textbf{a}_i}$ and $\dddot{\textbf{a}_i}$ are the first, second and third-order time derivatives of acceleration obtained from the previous time-step. $\gamma$ is an accuracy factor of order $\sim0.0001$. The acceleration $\textbf{a}_{i}$ on each body due to the gravity from all other bodies in the simulation is calculated using
\begin{equation}
\label{eqn:accel}
\textbf{a}^{n}_{i} = G \sum_{j=1}^{N} \frac{\textbf{r}_{ij}}{| \textbf{r}_{ij}^{3}|},
\end{equation}
where 
\begin{equation}
\textbf{r}_{ij} = \textbf{r}_{i} - \textbf{r}_{j},
\end{equation}
\begin{equation}
\textbf{v}_{ij} = \textbf{v}_{i} - \textbf{v}_{j},
\end{equation}
and G is the gravitational constant. The first-order time derivative of acceleration (often referred to as jerk) is given by
\begin{equation}
\label{eqn:jerk}
    \dot{\textbf{a}^{n}_{i}} = G \sum^{N}_{j=1} m_{j} \frac{\textbf{v}_{ij}}{| \textbf{r}_{ij}^{3}|} + \frac{3(\textbf{r}_{ij}\cdot \textbf{v}_{ij})\textbf{r}_{ij}}{| \textbf{r}_{ij}^{5}|}.
\end{equation}
The values for the positions and velocities of body $i$ are predicted at the end of the time-step
\begin{equation}
\textbf{r}^{n+1}_{i} = \textbf{r}^{n+1}_{i} + \textbf{v}^{n}_{i} \Delta t + \frac{1}{2} \textbf{a}^{n}_{i} \Delta t^{2} + \frac{1}{6} \dot{{\textbf{a}^{n}_{i}}} \Delta t^{3},
\end{equation}
\begin{equation}
\textbf{v}^{n+1}_{i} = \textbf{v}^{n+1}_{i} + \textbf{a}^{n}_{i} \Delta t + \frac{1}{2} \dot{\textbf{a}^{n}_{i}} \Delta t^{2}.
\end{equation}
The acceleration and jerk are once again calculated using (\ref{eqn:accel}) and (\ref{eqn:jerk}) respectively, using the the new positions and velocities. The second and third order time derivatives are then calculated at the start of the time-step
\begin{equation}
\ddot{\textbf{a}^{n}_{i}} = \frac{2[-3(\textbf{a}^{n}_{i} - \textbf{a}^{n+1}_{i}) - (2\dot{\textbf{a}^{n}_{i}} + \textbf{a}^{n+1}_{i})\Delta t]}{\Delta t^{2}},    
\end{equation}
\begin{equation}
\dddot{\textbf{a}^{n}_{i}} = \frac{6[2(\textbf{a}^{n}_{i} - \textbf{a}^{n+1}_{i}) + (\dot{\textbf{a}^{n}_{i}} + \textbf{a}^{n+1}_{i})\Delta t]}{\Delta t^{3}}.   
\end{equation}
The higher order terms are used to correct the position and velocity of the body, i.e.
\begin{equation}
\textbf{r}_{i}^{n+1} = \textbf{r}_{i}^{n} + \frac{1}{24} \ddot{\textbf{a}_{i}^{n}}\Delta t^{4} + \frac{1}{120} \dddot{\textbf{a}_{i}^{n}}\Delta t^{5},
\end{equation}
\begin{equation}
\textbf{v}_{i}^{n+1} = \textbf{v}_{i}^{n} + \frac{1}{6} \ddot{\textbf{a}_{i}^{n}}\Delta t^{3} + \frac{1}{24} \dddot{\textbf{a}_{i}^{n}}\Delta t^{4}.
\end{equation}

\section{Initial Conditions}
\begin{figure}
	\includegraphics[width=\columnwidth]{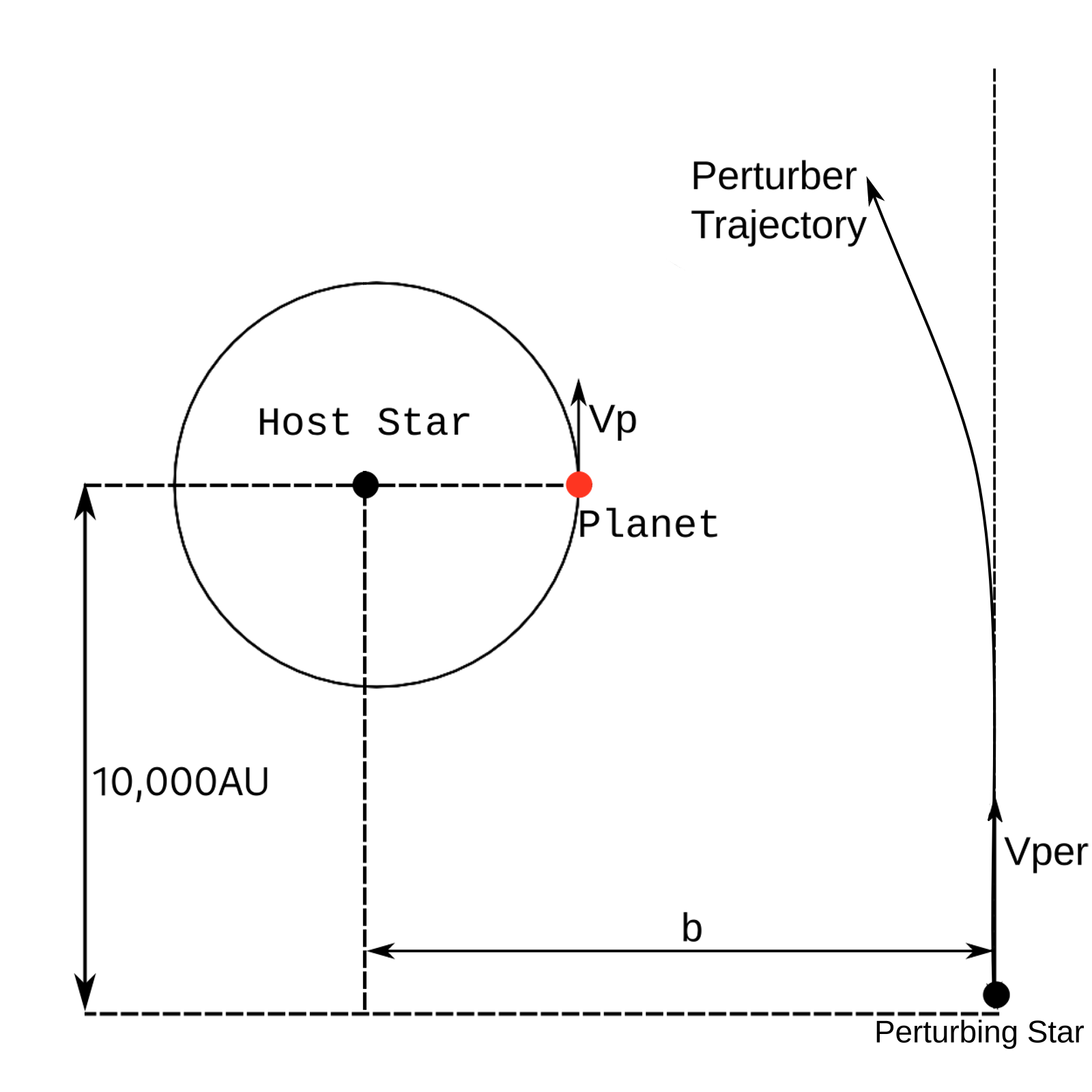}
    \caption{Diagram of the planet initially on a wide, circular orbit around its host. The path of the perturbing star (for a prograde approach), its velocity, $v_{per}$, and impact parameter, $b$, are shown.}
    \label{fig:diagram}
\end{figure}

We consider a star with mass $M_*=1\msol$ hosting a planet with mass $M_{\rm p}=1\jup$, on a wide, circular ($e=0$), Keplerian orbit with semi-major axis $a_p=100~\rm AU$. The planet is given a random true anomaly $f$, where $0\leq f<2\pi{}$, assuming a uniform distribution. We place a perturbing star with mass $M_{\rm per}$ at $x=b$, $y=\pm10,000~\rm AU$ relative to the centre of the mass of the star-planet system (see Fig. \ref{fig:diagram}), where $\rm~b$ is the impact parameter of the perturber. The perturbing star is given a velocity $v_{\rm per}$ that is initially parallel to the y-axis towards the star-planet system (the y-component of the perturber velocity is positive for a perturber at $y=-10000\rm~AU$ and negative for a perturber at $y=10000~{\rm AU}$). {The initial velocities of $1\rm~kms^{-1}$ and $3\rm~kms^{-1}$ of the perturbing star are informed from the distribution of velocities for close stellar encounters as seen in clusters \citep{Furesz:2008, Rochau:2010}. These are comparable to the velocity dispersions for young clusters such as the Orion Nebula Cluster (ONC) and NGC3603, which have typical densities of the order $10^{3}-10^{4}\rm~stars \, pc^{-3}$ \citep[e.g.][]{van-Altena:1988, Hillenbrand:1998, Proszkow:2009}. For a cluster of uniform density with a velocity dispersion $\sigma_{v}=4\rm~kms^{-1}$ and a stellar density of $10^{4}\rm~stars \, pc^{-3}$, \cite{Winter:2018} find that all stars have at least one encounter within $1000\rm~AU$ in the first $3\rm~Myr$ of their lives. That means that every star is expected to experience at least one encounter like the ones we simulate here (i.e. within our upper limit of 1000~AU for the perturber impact parameter). \cite{Bressert:2010} find that $<26\%$ of Young Stellar Objects are formed in environments where they are likely to interact with neighbouring stars.}

We note that the distance of closest approach between the perturber and the host star is smaller than the impact parameter of the perturber, as its path is bent towards the host star due to the gravitational interaction with the star-planet system.

We simulate the evolution of the planetary system as it interacts with the perturbing star for $250~\rm kyr$, varying the host mass and the perturber mass, initial y-component of velocity, direction of approach and impact parameter (see Table \ref{tab:ICs} for the parameter space explored). We perform 100 simulations for each combination of free parameters, choosing a random true anomaly each time, i.e. a total of 9600 simulations. We investigate the ejection rate and the orbital properties of the planets that remain bound to their host star post-encounter.

\begin{table}
	\centering
	\caption{Parameter space of initial conditions of the planetary system (see text for definition of the parameters).}
	\label{tab:ICs}
	\begin{tabular}{lc}
		\hline
		Parameter & Values\\
		\hline
		$M_{*}$ & $0.2,1,1.5 \rm \msol$\\
		$M_{\rm p}$ & $1 \rm \jup$\\
		$a_{\rm p}$ & $100 \rm~AU$\\
		$f$ & $0\leq f <2\pi$\\
		$e$ & $0$\\
		$b$ & $200,400,800,1200\rm AU$\\
		$M_{\rm per}$ & $0.1,1 \rm \msol$\\
		$v_{\rm per}$ ($v_{y}$) & $1,3 \ \rm km\ s^{-1}$\\
		Perturber approach direction & prograde, retrograde\\
		\hline
	\end{tabular}
\end{table}

\section{The Dynamical Stability of Planets On Wide Orbits}

Initially the giant planet is stable on a wide orbit ($a_{p}=100\rm AU$). The perturbing star interacts gravitationally with the planet-host system as it passes; in most cases the orbit of the planet is perturbed, altering its eccentricity and semi-major axis. In some cases the planet is scattered inwards towards the host star, whilst others are scattered outwards to much wider orbits. Some planets experience a strong enough interaction with the passing star that they are gravitationally liberated from their host, becoming free-floating planets. The perturbing star also interacts gravitationally with the host star, altering its path from a linear flyby to a parabolic flyby. 

\begin{figure}
	\includegraphics[width=\columnwidth]{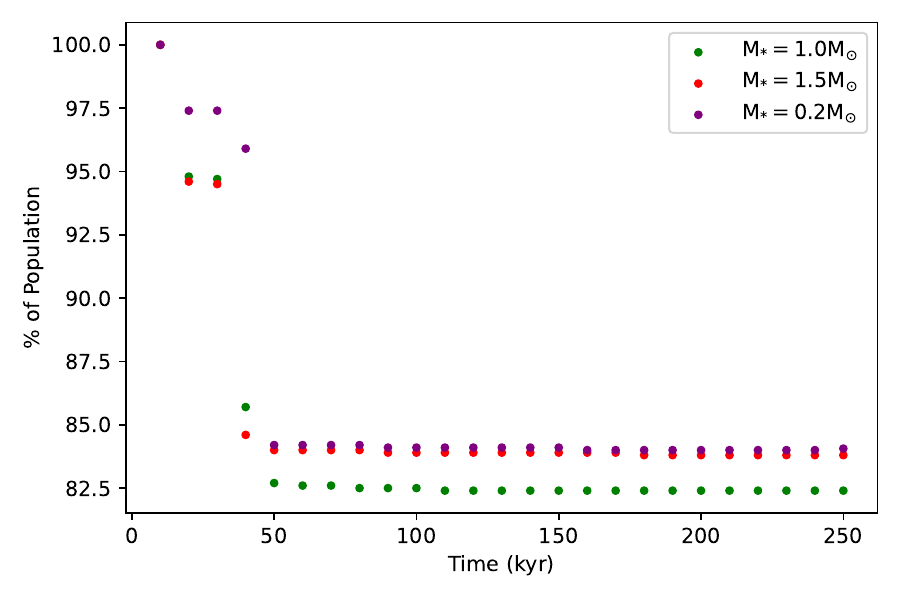}
    \caption{Percentage of planetary systems in which the wide-orbit planet remains bound to to the host star with respect to time. The statistical error for each point is ±2\%. There are no ejections after $\sim 50$~kyr.}
    \label{fig:overTime}
\end{figure}

The results of the simulations for different encounter parameters are summarised in Table \ref{tab:large} and in Table \ref{tab:general}. We extract four key statistics post-encounter to consider the effects of the perturbing star on the planetary system: (i) planet ejection rate, (ii) percentage of planets that scattered inwards, (iii) percentage of planets that scattered outwards, and (iv) percentage of planets with excited eccentricities ($e_{\rm f}>0.1$). We find that most ejections occur within the first $50\rm~kyr$ (see Fig. \ref{fig:overTime}).

\begin{table*}
	\centering
	\caption{Ejection rate, planet scattering pattern and planets with final eccentricities $e_{\rm f} > 0.1$ across 9600 simulations. $ M_{*}$ is the mass of the host star, $ M_{\rm per}$ is the mass of the perturber star, $\rm~b$ its impact parameter , $v_{\rm per}$ its initial velocity (along the y axis), $b_{\rm min}$ is the closest approach of the perturber to the star-planet system. "Inwards" and "Outwards" describe the percentage of planets that experience significant inwards and outwards scattering, respectively. Statistical errors are also quoted. Note that the percentages in the last 3 columns are calculated excluding the ejected planets.}
	\label{tab:large}
	\begin{tabular}{ccccccccc} 
		\hline
        $M_{*}$ & $M_{\rm per}$ ($\rm \msol$) & $~b$ (AU) & $v_{\rm per}$ ($\rm km s^{-1}$) & $~b_{\rm min}$ (AU) & Ejection Rate (\%) & Inwards (\%) & Outwards (\%) & $e_{\rm f}>0.1$\\
        \hline

        0.2 & 0.1 & 200 & 1 & 68 & 59$\pm5$ & 16$\pm3$ & 15$\pm3$ & 41$\pm4$\\
        0.2 & 1.0 & 200 & 1 & 25 & 95$\pm7$ & 0 & 5$\pm2$ & 5$\pm5$\\
        0.2 & 0.1 & 200 & 3 & 173 & 0 & 0 & 13$\pm3$ & 35$\pm4$\\
        0.2 & 1.0 & 200 & 3 & 119 & 42$\pm5$ & 11$\pm2$ & 44$\pm5$ & 59$\pm6$\\
        0.2 & 0.1 & 400 & 1 & 216 & 0 & 0 & 0 & 0\\
        0.2 & 1.0 & 400 & 1 & 74 & 60$\pm6$ & 0 & 41$\pm5$ & 41$\pm5$\\
        0.2 & 0.1 & 400 & 3 & 373 & 0 & 0 & 0 & 0\\
        0.2 & 1.0 & 400 & 3 & 303 & 0 & 0 & 7$\pm2$ & 43$\pm5$\\
        0.2 & 0.1 & 800 & 1 & 587  & 0 & 0 & 0 & 0\\
        0.2 & 1.0 & 800 & 1 & 273 & 0 & 0 & 12$\pm2$ & 63$\pm6$\\
        0.2 & 0.1 & 800 & 3 & 773 & 0 & 0 & 0 & 0\\
        0.2 & 1.0 & 800 & 3 & 697 & 0 & 0 & 0 & 0\\
        0.2 & 0.1 & 1200 & 1 & 983 & 0 & 0 & 0 & 0\\
        0.2 & 1.0 & 1200 & 1 & 560 & 0 & 0 & 0 & 0\\
        0.2 & 0.1 & 1200 & 3 & 1174 & 0 & 0 & 0 & 0\\
        0.2 & 1.0 & 1200 & 3 & 1099 & 0 & 0 & 0 & 0\\
        \hline
        1.0 & 0.1 & 200 & 1 & 23 & 18$\pm3$ & 35$\pm4$ & 42$\pm5$ & 76$\pm6$\\
        1.0 & 1.0 & 200 & 1 & 12 & 76$\pm6$ & 3$\pm1$ & 19$\pm3$ & 22$\pm3$\\
        1.0 & 0.1 & 200 & 3 & 119 & 5$\pm2$ & 22$\pm3$ & 27$\pm4$ & 56$\pm5$\\
        1.0 & 1.0 & 200 & 3 & 84 & 70$\pm6$ & 12$\pm2$ & 14$\pm3$ & 30$\pm4$\\
        1.0 & 0.1 & 400 & 1 & 79 & 17$\pm3$ & 24$\pm3$ & 46$\pm5$ & 67$\pm6$\\
        1.0 & 1.0 & 400 & 1 & 45 & 62$\pm6$ & 12$\pm2$ & 26$\pm4$ & 38$\pm4$\\
        1.0 & 0.1 & 400 & 3 & 308 & 0 & 0 & 0 & 0\\
        1.0 & 1.0 & 400 & 3 & 251 & 10$\pm2$ & 19$\pm3$ & 16$\pm3$ & 44$\pm5$\\
        1.0 & 0.1 & 800 & 1 & 292 & 0 & 0 & 0 & 0\\
        1.0 & 1.0 & 800 & 1 & 175 & 25$\pm4$ & 21$\pm3$ & 25$\pm4$ & 65$\pm6$\\
        1.0 & 0.1 & 800 & 3 & 705 & 0 & 0 & 0 & 0\\
        1.0 & 1.0 & 800 & 3 & 636 & 0 & 0 & 0 & 0\\
        1.0 & 0.1 & 1200 & 1 & 592 & 0 & 0 & 0 & 0\\
        1.0 & 1.0 & 1200 & 1 & 379 & 0 & 0 & 0 & 22$\pm3$\\
        1.0 & 0.1 & 1200 & 3 & 1107 & 0 & 0 & 0 & 0\\
        1.0 & 1.0 & 1200 & 3 & 1036 & 0 & 0 & 0 & 0\\
        \hline
		1.5 & 0.1 & 200 & 1 & 45 & 8$\pm2$ & 39$\pm4$ & 42$\pm5$ & 76$\pm6$\\
        1.5 & 1.0 & 200 & 1 & 65 & 63$\pm6$ & 9$\pm2$ & 23$\pm3$ & 35$\pm4$\\
        1.5 & 0.1 & 200 & 3 & 98 & 21$\pm3$ & 17$\pm3$ & 33$\pm4$ & 42$\pm5$\\
        1.5 & 1.0 & 200 & 3 & 72 & 56$\pm5$ & 19$\pm3$ & 21$\pm3$ & 45$\pm5$\\
        1.5 & 0.1 & 400 & 1 & 71 & 10$\pm2$ & 32$\pm4$ & 40$\pm4$ & 76$\pm6$\\
        1.5 & 1.0 & 400 & 1 & 50 & 63$\pm6$ & 12$\pm2$ & 18$\pm3$ & 37$\pm4$\\
        1.5 & 0.1 & 400 & 3 & 276 & 0 & 0 & 0 & 0\\
        1.5 & 1.0 & 400 & 3 & 228 & 11$\pm2$ & 12$\pm2$ & 23$\pm3$ & 49$\pm5$\\
        1.5 & 0.1 & 800 & 1 & 214 & 0 & 0 & 0 & 10$\pm2$\\
        1.5 & 1.0 & 800 & 1 & 142 & 0 & 13$\pm3$ & 31$\pm4$ & 62$\pm6$\\
        1.5 & 0.1 & 800 & 3 & 666 & 0 & 0 & 0 & 0\\
        1.5 & 1.0 & 800 & 3 & 601 & 0 & 0 & 0 & 0\\
        1.5 & 0.1 & 1200 & 1 & 455 & 0 & 0 & 0 & 0\\
        1.5 & 1.0 & 1200 & 1 & 313 & 0 & 4$\pm1$ & 0 & 50$\pm5$\\
        1.5 & 0.1 & 1200 & 3 & 1067 & 0 & 0 & 0 & 0\\
        1.5 & 1.0 & 1200 & 3 & 999 & 0 & 0 & 0 & 0\\
		\hline
	\end{tabular}
\end{table*}

\subsection{Planet ejection}
A planet has been ejected (i.e. has become a free-floating planet) when the planet-host pair have a binding energy $\rm E_{b}>0$ post-encounter. The binding energy is calculated using:

\begin{equation}
    E_{\rm b} = \frac{\mu v^{2}}{2} - \frac{GM_{*} M_{\rm p}}{r},
\end{equation}
where
\begin{equation}
    \mu = \frac{M_{*} M_{\rm p}}{M_{*}+ M_{\rm p}},
\end{equation}
$v$ is the velocity of the planet relative to the host star, and $r$ its distance from it. We find that across the total population of simulated wide-orbit planets, close encounters with stars incite an ejection in $\sim18\%$ of cases. There is a significant decrease in ejection rate for a $0.1\msol$ flyby with an impact parameter $b \geq 800\rm~AU$. Moreover, the interaction between the perturbing star and planet is not significant enough to unbind the planet from its host star for encounters with an impact parameter $b \geq 1200\rm~AU$. We find that the ejection rate is independent of the host mass (see Table~2; we observe no ejections for planets orbiting $0.1\msol$ and $1\msol$ hosts, and an ejection rate of $\sim4\%$ for $1.5\msol$ hosts), but closely related to the maximum impact parameter that can incite a significant perturbation of the orbit of a long-period giant planet. A higher host mass directly correlates to a stronger dynamical interaction with the passing star, altering its path and leading to an encounter up to $\sim75\%$ closer than the initial impact parameter (see Table \ref{tab:large}). A lower mass host interacts more weakly with the perturbing star but has a weaker gravitational pull to the planet, leaving the planet more prone to significant perturbations from close-in encounters. Therefore, the statistics of observed giant planets around different mass stars \citep{Galicher2016,Vigan:2017a,Vigan:2021w} are not expected to be skewed due to interactions from passing stars.
\subsection{Inwards/outwards Scattering of Wide-Orbit Planets}

In cases where the planet is not ejected, the orbit of the planet may still be perturbed post-encounter. We consider that the orbit of a planet is perturbed when its semi-major axis changes by $\Delta a_{p} \geq5\rm AU$, whilst also remaining bound to their host ($E_{\rm b}<0$). Therefore, interactions with a $\Delta a_{p}<5\rm~AU$ are considered not to constitute a significant change in the architecture of the star-planet system. We find that encounters with slow ($v_{\rm per}=\rm 1kms^{-1}$), massive ($M_{\rm per}=~1\msol$) stars significantly perturb the orbit of wide-orbit giants even with impact parameter $\geq 800~\rm AU$ where gravitational interactions between the planet and the perturbing star are too weak to unbind the planet from its orbit around its host star (see Table \ref{tab:large}). Wide-orbit planets that have had their orbits significantly perturbed may be less stable, and as a result could be prone to ejection by subsequent interactions. We observe a similar distribution across different host masses for both the distribution of semi-major axes and eccentricities of the planet post-encounter (see Fig. \ref{fig:sm_0_200_1}, \ref{fig:ecc_1}).

\subsection{The Effect of the Host Star Mass}

\subsubsection{Flybys Around a 1-$M_{\odot}$ Host Star}

\begin{table}
    \centering
  \caption{Overview of properties of wide-orbit planets. $E_{\rm b}$ is the binding energy of the star-planet system,  $a_{\rm p,f}$ the final semi-major axis of the planet, and $e_{\rm f}$ its final eccentricity. Note that percentages of bound planets (denoted by a $*$ superscript) are calculated excluding ejected planets.}
      \label{tab:general}
    \begin{tabular}{cccc}
    \hline
    $\rm M_{*}$ & Parameter & Criteria & \% of Population\\
    \hline
    0.2 & Bound & $E_{\rm b}<0$ & 84 $\pm2$\\
    0.2 & Unbound & $E_{\rm b}>0$ & 16 $\pm1$\\
    0.2 & Perturbed$^*$ & $|\Delta E_{\rm b}|>5\%$ & 12$\pm1$\\
    0.2 & Inwards$^*$ & $a_{\rm p,f}<95\rm~AU$ & 2$\pm1$\\
    0.2 & Outwards$^*$ & $a_{\rm p,f}>105\rm~AU$ & 8$\pm1$\\
    0.2 & Eccentric$^*$ & $e_{\rm f}>0.1$ & 18$\pm1$\\
        \hline
            1.0 & Bound & $E_{\rm b}<0$ & 82$\pm2$\\
    1.0 & Unbound & $E_{\rm b}>0$ & 18$\pm1$\\
    1.0 & Perturbed$^*$ & $|\Delta E_{\rm b}|>5\%$ & 27$\pm1$\\
    1.0 & Inwards$^*$ & $a_{\rm p,f}<95$AU & 9$\pm1$\\
    1.0 & Outwards$^*$ & $a_{\rm p,f}>105\rm~AU$ & 13$\pm1$\\
    1.0 & Eccentric$^*$ & $e_{\rm f}>0.1$ & 26$\pm1$\\
        \hline
            1.5 & Bound & $E_{\rm b}<0$ & 84$\pm1$\\
    1.5 & Unbound & $E_{\rm b}>0$ & 16$\pm1$\\
    1.5 & Perturbed$^*$ & $|\Delta E_{\rm b}|>5\%$ & 29$\pm1$\\
    1.5 & Inwards$^*$ & $a_{\rm p,f}<95\rm~AU$ & 10$\pm1$\\
    1.5 & Outwards$^*$ & $a_{\rm p,f}>105\rm~AU$ & 14$\pm1$\\
    1.5 & Eccentric$^*$ & $e_{\rm f}>0.1$ & 30$\pm1$\\
    \hline
    \end{tabular}
\end{table}

\begin{figure}
	\includegraphics[width=\columnwidth]{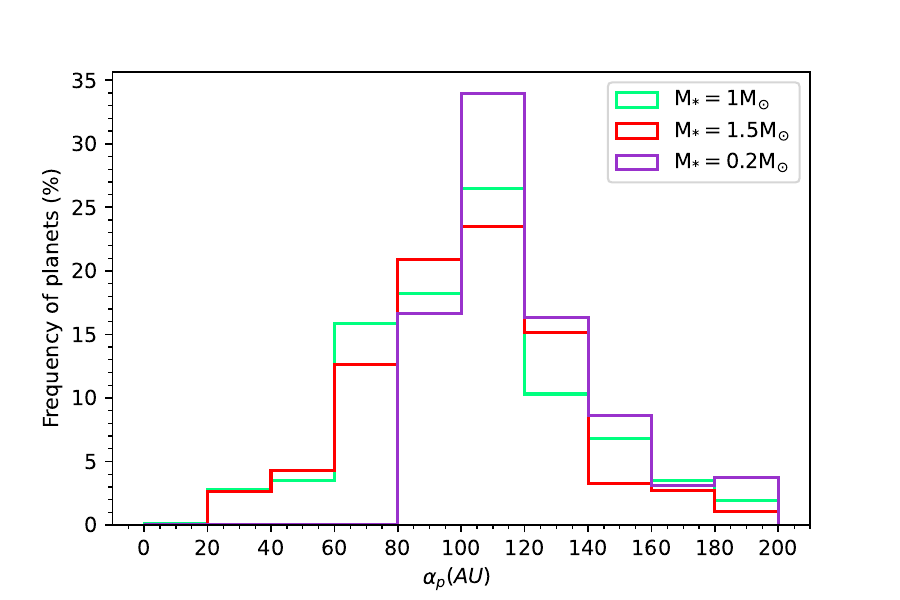}
    \caption{Distribution of semi-major axes post-encounter for bound planets of semi-major axis $\geq\pm5$~AU from the initial semi-major axis ($100$~AU). The errors are  of the order of $\pm$1-2$\%$} 
    \label{fig:sm_0_200_1}
    
    \includegraphics[width=\columnwidth]{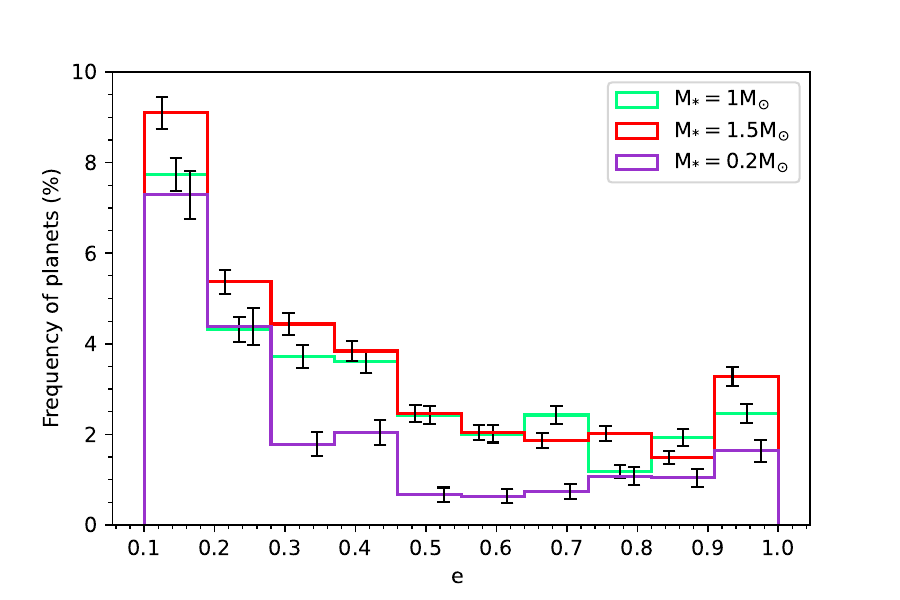}
    \caption{Distribution of eccentricities of all planets still bound post encounter orbiting host stars of mass $0.2\msol$,$1\msol$ and $1.5\msol$.}
    \label{fig:ecc_1}
    
    \includegraphics[width=\columnwidth]{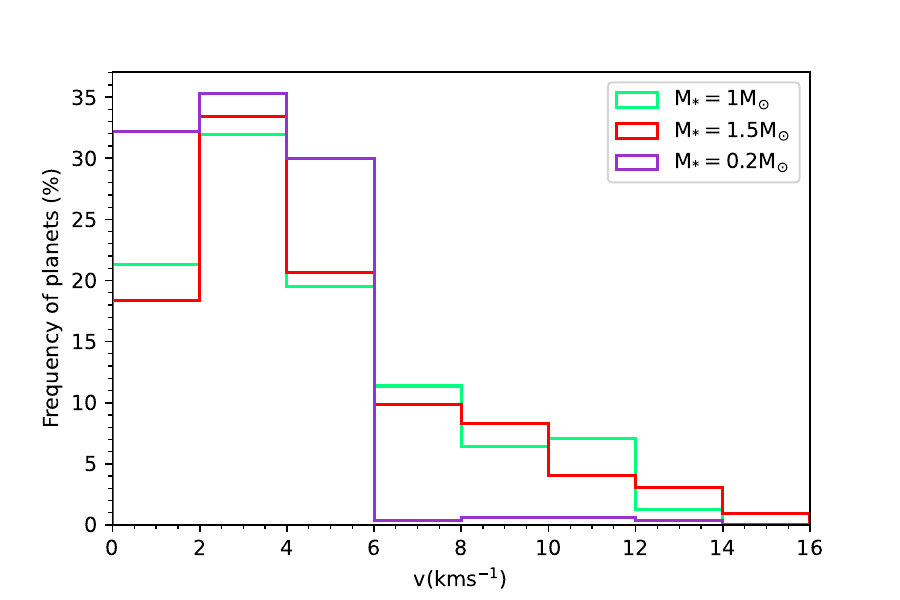}
    \caption{Distribution of velocities of the unbound population of planets formed as a result of ejections due to simulated encounters. The errors are of the order of $\pm$1-2$\%$} 
    \label{fig:vf_0_100_1}
\end{figure}

Across the entire parameter space for the properties of the perturbing star, $\sim78\%$ of the bound planets orbiting a $1\msol$ host are found to remain within $\pm5$~AU of their initial semi-major axis, i.e. their orbits are not significantly perturbed (see Table \ref{tab:general}).  This unperturbed case is predominant in encounters with a $M_{\rm per}=0.1\msol$ perturber, and encounters with an impact parameter $b\geq800$~AU. Fig. \ref{fig:sm_0_200_1} (green line) shows the distribution of bound planets with perturbed orbits and semi-major axes within $0-200~\rm AU$.

In Fig. \ref{fig:ecc_1} we plot the distribution of the final eccentricities of the planets (green line corresponds to the 1-$\msol$ host star case). This figures shows that close encounters may leave wide-orbit planets with excited eccentricities. Across 3200 simulated encounters with a giant planet orbiting a $1\msol$ host, $\sim~26\%$ wide-orbit giants were observed with eccentricities $e\geq0.1$ post-encounter. Of the eccentric population, $\sim53\%$ of planets have low eccentricities ($0.1\leq e < 0.4$). $\sim32\%$ have higher eccentricities $0.4\leq e < 0.8$, with the remaining $\sim15\%$ having extreme eccentricities $0.8\leq e < 1$. We find that it is possible for the eccentricity of the planet's orbit to increase whilst its semi-major axis remains almost unaffected (see Table \ref{tab:general}). This behaviour has also been observed by \cite{Parker:2012}. Planets on eccentric orbits may be less stable, and highly eccentric planets on wide-orbits could be particularly prone to ejection as a result of further perturbation through factors such as subsequent flybys or, in the case of multi-planet systems, planet-planet scattering \citep{Veras:2012,Li:2015b,Li:2016k}.

We present the distribution of velocities for giant planets ejected from their system post-encounter in Fig. \ref{fig:vf_0_100_1}. We find typical ejection velocities of a few ${\rm km\ s^{-1}}$ in agreement with previous works \citep{Parker:2012}.

\subsubsection{Flybys Around a 0.2-$\msol$ Star}
We find that, across 3200 simulations, $\sim~16\%$ of giant planets orbiting a $0.2~\msol$ host are ejected from their system due to strong gravitational interactions with a perturbing star. The results of the simulations for encounters with a $0.2\rm~M_{\odot}$ host star are summarised in Table \ref{tab:general}.

We find $\sim90\%$ of giant planets orbiting a $0.2~\msol$ host star post-encounter remain within $\pm~5$~AU of their initial semi-major axis. The proportion of planets perturbed due to a perturbing star are lower when compared to planets orbiting higher mass stars. Encounters with a flyby with impact parameter $\leq400\rm~AU$ and velocity $v_{\rm per}=1\rm~kms^{-1}$ scatter the planet most significantly; most of the planets are scattered outwards and excited to a more eccentric orbit ($e\geq0.1$) (see Table \ref{tab:large}). Fewer planets orbiting $0.2~\msol$ experience inwards scattering compared to giant planets orbiting higher mass hosts (see Fig. \ref{fig:sm_0_200_1}). We find that it is more common for wide-orbit giant planets orbiting low-mass stars to get scattered outwards post-encounter with a perturbing star than wide-orbit giants orbiting more massive hosts.

$\sim18\%$ of planets orbiting a $0.2\msol$ host star are excited to an orbit with eccentricity $e\geq0.1$. We find a similar distribution of eccentricities for the planets still bound to a $0.2\msol$ host post encounter with a perturbing star compared to planets orbiting higher mass hosts (see Fig. \ref{fig:ecc_1}). Of the planets with eccentricity $e\geq0.1$ post-encounter, $\sim66\%$ are found to have an orbital eccentricity of $0.1\leq~e\leq0.4$. We find $\sim21\%$ of the eccentric population orbiting a $0.2\msol$ host with eccentricity $0.4\leq~e\leq0.8$, and the remaining $\sim13\%$ on highly eccentric orbits ($e\geq0.8$).

Planets ejected from their orbit around $0.2\msol$ are found to be ejected with a narrower distribution of ejection velocities than those around a $1\msol$ star; $\sim87\%$ of the ejected planets have velocities $\leq5\rm~kms^{-1}$.

\subsubsection{Flybys Around a 1.5-$\msol$ Star}
We find that, across 3200 simulations, $\sim16\%$ of giant planets orbiting a $1.5~\msol$ host are liberated from their orbit due to strong gravitational interactions with a perturbing star. The results of the simulations for encounters with a $1.5\rm~\msol$ host star are summarised in Table \ref{tab:general}.

We find $\sim78\%$ of giant planets orbiting a $1.5~\msol$ host to be within $\pm~5$~AU of their initial semi-major axis after the interaction, similar to the proportion of planets orbiting a $1~\msol$ host. We find planets orbiting a $1.5\msol$ host star to be most affected by interactions with perturbing stars of mass $M_{\rm per}=1\msol$, with velocity $v_{\rm per}=1\rm~kms^{-1}$.

We expect a greater proportion of $1.5\msol$ stars to host giant planets on extremely eccentric orbits in comparison to giant planets orbiting lower mass hosts. Of the $\sim30\%$ planets found on eccentric ($e\geq0.1$) orbits post-encounter, $\sim58\%$ are found on orbits with eccentricities $0.1\leq~e\leq0.4$. $\sim28\%$ of the population are found on orbits with eccentricity $0.4\leq~e\leq0.8$; the remaining $\sim14\%$ of planets bound to a $1.5\msol$ host are found on orbits with extreme eccentricities ($e\geq0.8$).

We find that the distribution of velocities for giant planets liberated  $1.5\msol$ hosts are quite similar to those liberated from a $1\msol$ star (see Fig.~\ref {fig:vf_0_100_1}).

\section{Discussion}
\begin{figure}
	\includegraphics[width=\columnwidth]{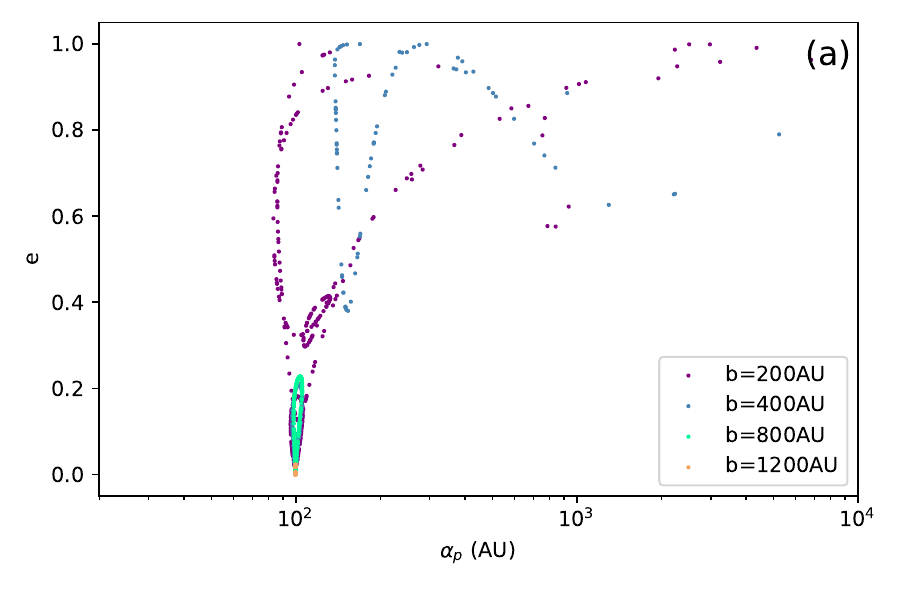}
	\includegraphics[width=\columnwidth]{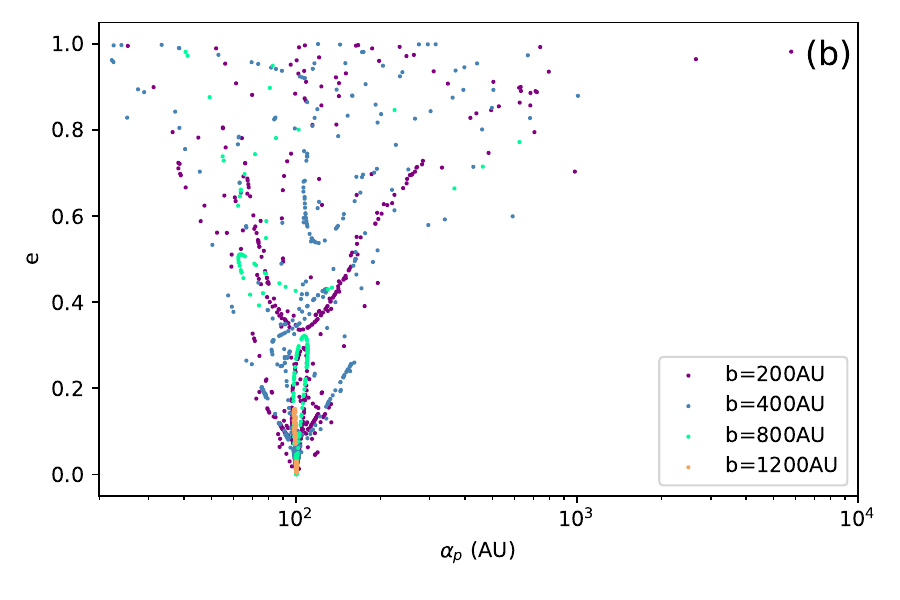}
	\includegraphics[width=\columnwidth]{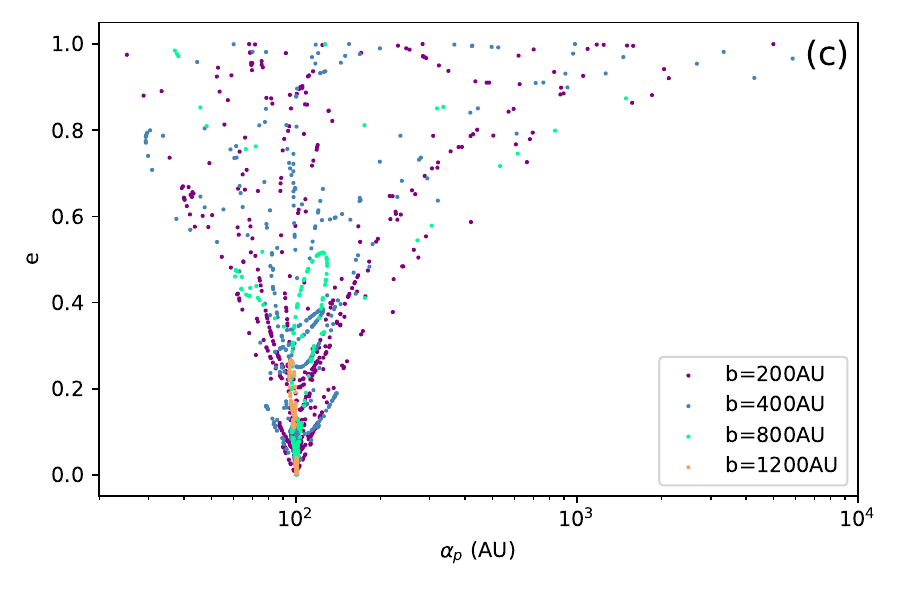}
    \caption{Eccentricity against semi-major axis of the bound population post-encounter, colour-mapped according to the impact parameter of the perturbing star. Each plot corresponds to the fate of planets orbiting host stars of mass a) $0.2\msol$, b) $1\msol$ and c) $1.5\msol$.}
    \label{fig:ecc_sm_b}
\end{figure}

\begin{figure}
	\includegraphics[width=\columnwidth]{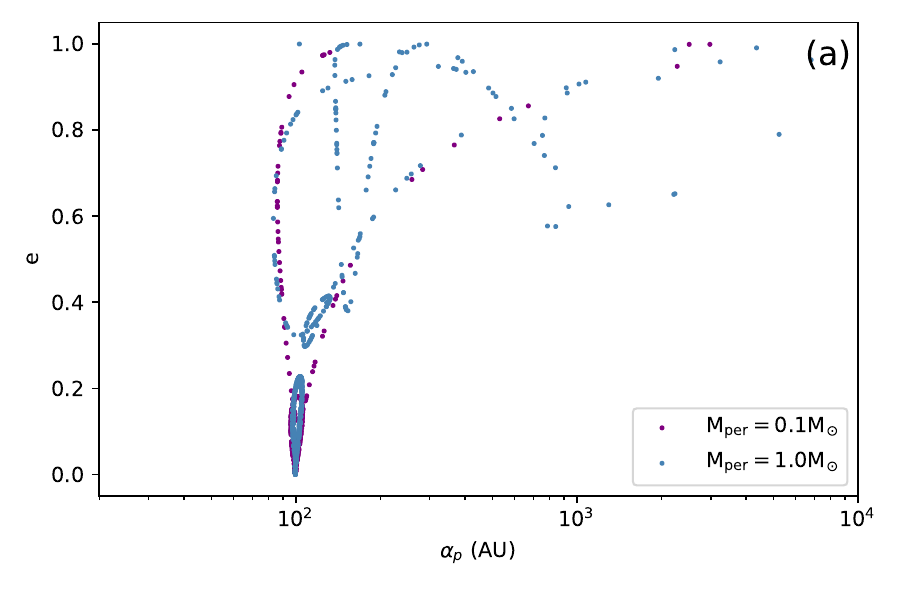}
	\includegraphics[width=\columnwidth]{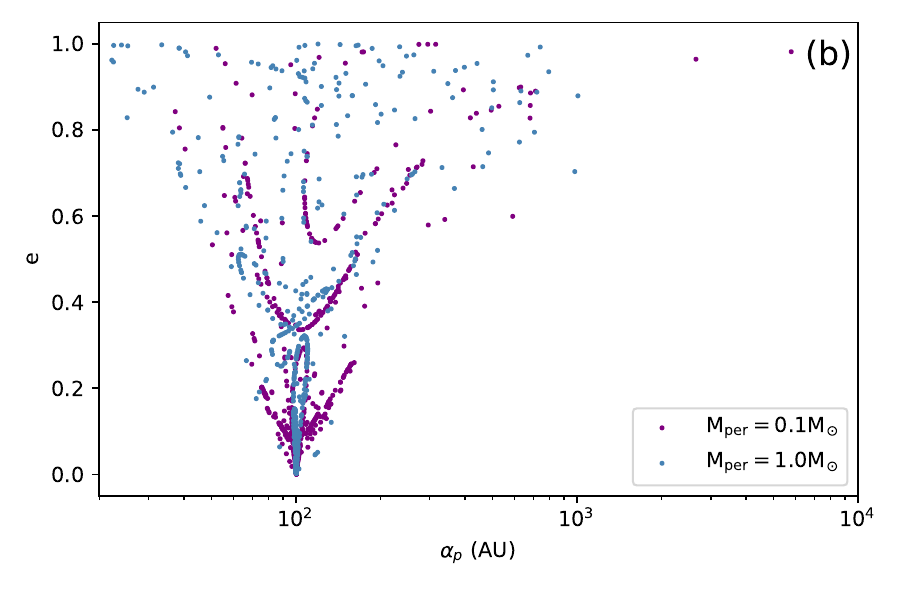}
	\includegraphics[width=\columnwidth]{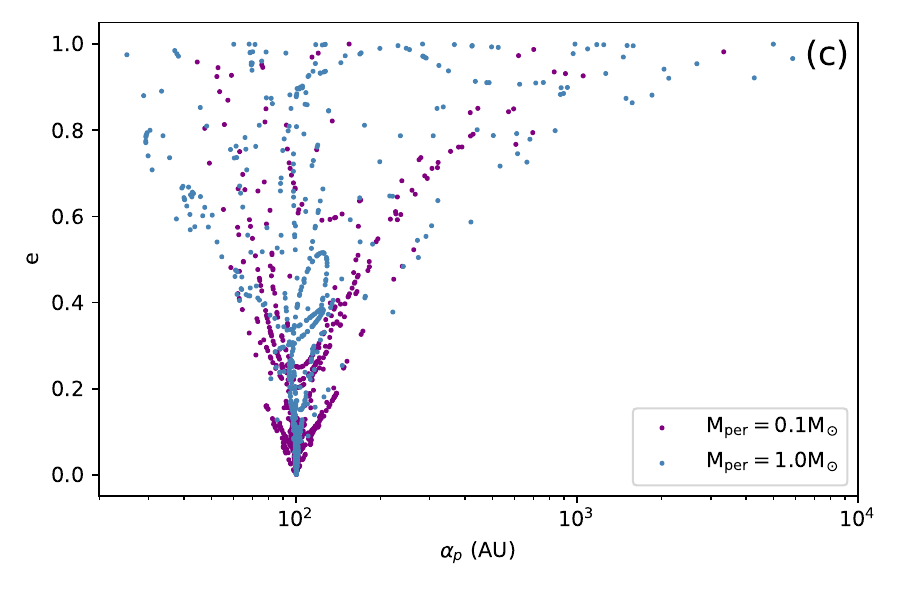}
    \caption{Eccentricity against semi-major axis of the bound population post-encounter, colour-mapped according to the mass of the perturbing star. Each plot corresponds to the fate of planets orbiting host stars of mass a) $0.2\msol$, b) $1\msol$ and c) $1.5\msol$.}
    \label{fig:ecc_sm_mass}
\end{figure}

\begin{figure}
	\includegraphics[width=\columnwidth]{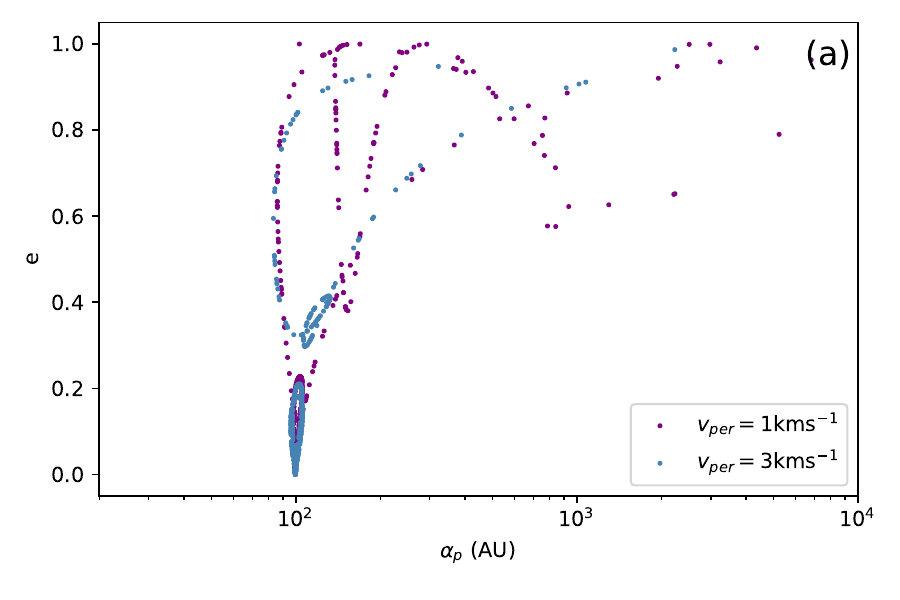}
	\includegraphics[width=\columnwidth]{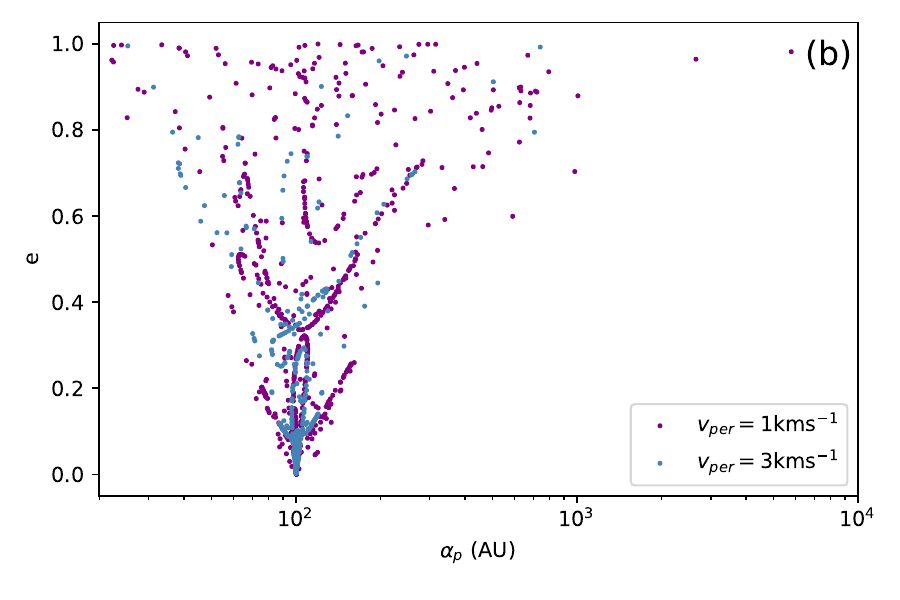}
	\includegraphics[width=\columnwidth]{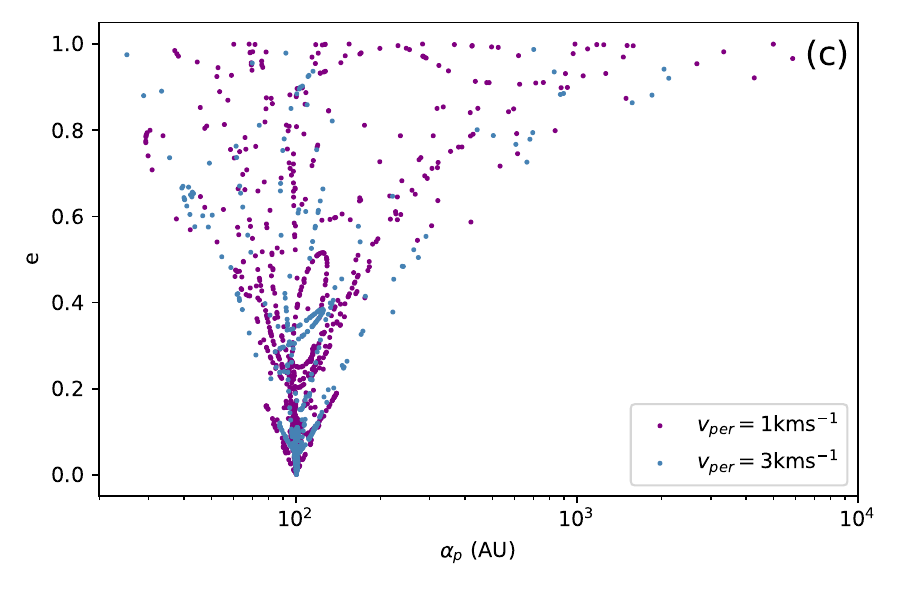}
    \caption{Eccentricity against semi-major axis of the bound population post-encounter, colour-mapped according to the velocity of the perturbing star. Each plot corresponds to the fate of planets orbiting host stars of mass a) $0.2\msol$, b) $1\msol$ and c) $1.5\msol$.}
    \label{fig:ecc_sm_vel}
\end{figure}

We present plots showing the distribution of the semi-major axes of the planets post-encounter with respect to eccentricity, colour-mapped with respect to each component of the parameter space for the perturbing star (see Fig. \ref{fig:ecc_sm_b}, \ref{fig:ecc_sm_mass}, \ref{fig:ecc_sm_vel}). The graph shows similarities to what is seen by \cite{Parker:2012}, who simulate the dynamical evolution of a $1\rm~M_{J}$ planet with semi major axis $5\rm~AU$ and $30\rm~AU$ in young sub structured star clusters. There is a significant population of perturbed planets that show increased eccentricity, with a subset of them scattered inwards or outwards. We observe a more extended distribution of eccentricities as the semi-major axis of the planet diverges from its initial semi-major axis prior to the encounter; this is due to the weaker binding energies between the planet and its host star as we simulate a planet on a significantly wider-orbit than in \cite{Parker:2012}. Further, we observe that a small number of planets may remain bound on extremely eccentric, ultra-wide $(a_{p}\geq1000\rm~AU)$ orbits. Such wide-orbit planets may correspond to  ultra-wide cold-Jupiters observed with the COCONUTS survey \citep{Zhang:2021}. The distribution of semi-major axes appear to be more extended with an increasing flyby mass and decreasing initial velocity and impact parameter. In the following sections we discuss the effect of each of the parameters varied in our runs.

\subsection{The Role of the Impact Parameter}
 We find that the extent of scattering a planet experiences depends heavily on the impact parameter (see Fig. \ref{fig:ecc_sm_b}). The smaller the impact parameter, the generally wider the separation between the planet and its host post encounter, given that the planet remains bound. Fig. \ref{fig:ecc_sm_b} shows little correlation between eccentricity and the impact parameter of the flyby. For $\rm~b=200,400,800\rm~AU$ the distribution is similar with a population of planets still bound with extreme eccentricities. The planet orbits have not been significantly altered for perturbers passing outside $\rm~b=1200\rm~AU$, placing an upper limit for the impact parameter capable of causing significant gravitational perturbations.

\subsection{The Role of the Perturber Mass}
We find that encounters with perturbing stars of mass $\rm M_{\rm per}=~1\msol$ scatter giant planets outwards more frequently than perturbing stars with mass $\rm M_{\rm per}=~0.1\msol$ (see Fig. \ref{fig:ecc_sm_mass}). Significant interactions occur more frequently between a wide-orbit giant planet and a $\rm M_{\rm per}=~1\msol$ perturbing star than with a star of lower mass. This is particularly evident with slow, close-in flybys where interactions between the perturbing star and the planet are the strongest (see Table \ref{tab:large}). Further, we observe that more massive perturbing stars cause a significantly greater number of ejections. For an encounter between planetary system with host mass $\rm~M_{*}=~0.2\msol$ and a perturbing star with $v_{\rm~per}=~1\rm~kms^{-1}$,$\rm~b_{\rm~per}=~200\rm~AU$, we find that $\sim~59\%$ of planets are ejected due to an encounter with a $0.1\msol$ perturbing star compared to $\sim95\%$ of planets ejected due to an encounter with a $1\msol$ perturbing star (see Table~\ref{tab:large}).

\subsection{The Role of the Perturber Velocity}
Fig. \ref{fig:ecc_sm_vel} shows a similar distribution of planets for both perturbing stars of velocity $v_{\rm~per}=1\rm~kms^{-1}$ and $v_{\rm~per}=3\rm~kms^{-1}$. We observe a higher frequency of ejections as a result of encounters with perturbing stars moving at lower velocities (see Table \ref{tab:large}). Encounters with a slower moving perturbing star allow for gravitational interaction between the planet and a perturbing star over a longer period of time, leading to more interactions sufficiently strong to incite an ejection.

\subsection{Host Mass}
We observe a similar distribution of semi-major axes and eccentricities of planets that remain bound post-encounter independent of the mass of host star. We find that the distribution of planets orbiting a low-mass ($\rm M_{*}=0.1\msol$) host star post-encounter present a narrower structure when compared to that of encounters involving a $1\msol$ host star (see Fig. \ref{fig:ecc_sm_b}, \ref{fig:ecc_sm_mass}, \ref{fig:ecc_sm_vel}). Giant planets orbiting low-mass hosts are significantly more susceptible to ejections by close ($\rm b=200~AU$), low-velocity encounters with a rapid reduction in ejection rate with increasing impact parameter. This is likely due to a low-mass star interacting more weakly with the perturbing star. Despite the planet having a weaker gravitational bond to its host, the path of perturbing star is left unperturbed leading to a overall weaker dynamical interaction with the planet. Therefore, we find similar ejection rates for giant planets orbiting hosts of different mass (see Table \ref{tab:large}). Moreover, we find a greater proportion of $1.5\msol$ stars to host giant planets on extremely eccentric orbits in comparison to giant planets orbiting lower mass hosts (see Fig \ref{fig:ecc_sm_b}a compared to \ref{fig:ecc_sm_b}b).



\section{Conclusions}
We have performed $N$-body simulations investigating the significance of close stellar flybys as a mechanism for perturbing young planetary systems hosting giant planets on wide orbits. We considered stars of mass $M_{*}=0.2,1,1.5\msol$ hosting a 1$\jup$ planet on a wide, circular orbit at $a=100\rm~AU$. A perturbing star passes by the star-planet system with a varying impact parameter and initial velocity, relative to the centre of the mass of the star-planet binary. We considered a total of 16 unique combinations of mass, impact parameter and velocity for the perturbing star on both prograde and retrograde approaches. We dynamically evolved 100 realisations per combination of parameters over a timescale of $250\rm~kyr$, leading to a total of 9600 simulated flybys. Of course, the parameter space investigated is only a small part for the variety of interactions that may happen in a cluster but the results are indicative of the general trends to be expected. 
The main results of our study are the following:

(i) The fraction of wide-orbit giant planets liberated from their host star as a result of dynamical interactions with a passing star is independent of the mass of the host star. Planets orbiting lower mass stars are more weakly gravitationally bound to their host, and hence more prone to strong interactions with a close-in perturbing star causing them to be liberated from their host star. However, interactions between the host star and the flyby also play a key role; dynamical interactions between the host and the flyby can reduce the closest approach of the flyby, leading to stronger interactions with the planet over a wider range of impact parameters. This is more prominent in simulations with a $1\msol$ or $1.5\msol$ host, where the orbital properties of the planet are still perturbed by flybys with greater impact parameter than the cutoff of $\sim800\rm~AU$ observed for encounters with a $0.2\msol$ host. As a result of the above competing influences, the ejection rates of wide orbit planets due to passing stars are similar for planets both around low-mass and high-mass stars. Therefore, we do not expect a dependence of the observed occurrence rates of wide orbit planets on the mass of the host star.

(ii) Stellar flybys may produce a population of giant planets on extremely wide, highly eccentric orbits. Simulated encounters with a $0.2\msol$ host show that a large portion of this population appears to have been liberated from the system. The percentage of high-eccentricity planets on wide orbits is found to be higher for higher mass host stars. Giant planets excited to extreme eccentricities $e\geq0.85$ may pass through the inner region of the planetary system, where terrestrial planets may exist \citep{Parker:2012}. Such interactions may have a significant impact on the dynamical evolution of the innermost planets of the system, potentially instigating ejections as a result of planet-planet scattering.

(iii) The extent of which the orbit a wide-orbit giant planet is perturbed as a result of dynamical interactions with a passing star is strongly correlated with the mass of the perturbing star and is inversely proportional to its velocity. Moreover, flybys with a lower impact parameter affect the orbit of the giant planet more significantly than flybys further away from the host-planet system. Encounters with a perturbing star with impact parameter $\geq800\rm~AU$ were seen to have a considerably weaker effect on the orbit of the planet.

Our results demonstrate that even one encounter of a planetary system containing a wide-orbit planet with a passing star in a cluster environment may have a significant effect on the survival of the planet on a wide-orbit, if the encounter is close enough ($\leq 1200\rm~AU$). For the specific set parameters investigated in this study, just one such encounter leads to an $\sim17\%$ chance of an ejection, and a $\sim 21\%$ chance of scattering to an eccentric orbit making it even more prone to future interactions. Moreover, there is a $\sim16\%$ probability of the wide-orbit planet getting scattered to an orbit with semi-major axis difference greater than 5~AU from the initial one. These percentages are even more significant if we consider encounters with impact parameter $\leq 800$~AU, as there are almost no change in the planetary orbits for encounters with impact parameter 1200~AU. Therefore, only a few encounters in a young star-forming environment \citep[e.g.][]{Rawiraswattana:2023i} are sufficient to eliminate the almost the entire initial population of wide-orbit planets. 

We conclude that the lack of a high occurrence rate of wide-orbit planets revealed by observational surveys does not exclude the possibility that such planetary systems may initially be abundant, and therefore the disc-instability model may be relevant to planetary formation.

\section*{Acknowledgements}
We would like to thank the referee Richard Parker for his constructive report that has improved the clarity of the paper. We thank D. Hubber for developing the $N$-body code that was used for this work. We acknowledge support from STFC grant ST/X508329/1.


\section*{Data Availability}
The data used for this paper can be provided by contacting the authors.


\bibliographystyle{mnras}
\bibliography{myref} 




\appendix
\section{Patterns on the the semi major axis-eccentricity graphs}

\begin{figure}
	\includegraphics[width=\columnwidth]{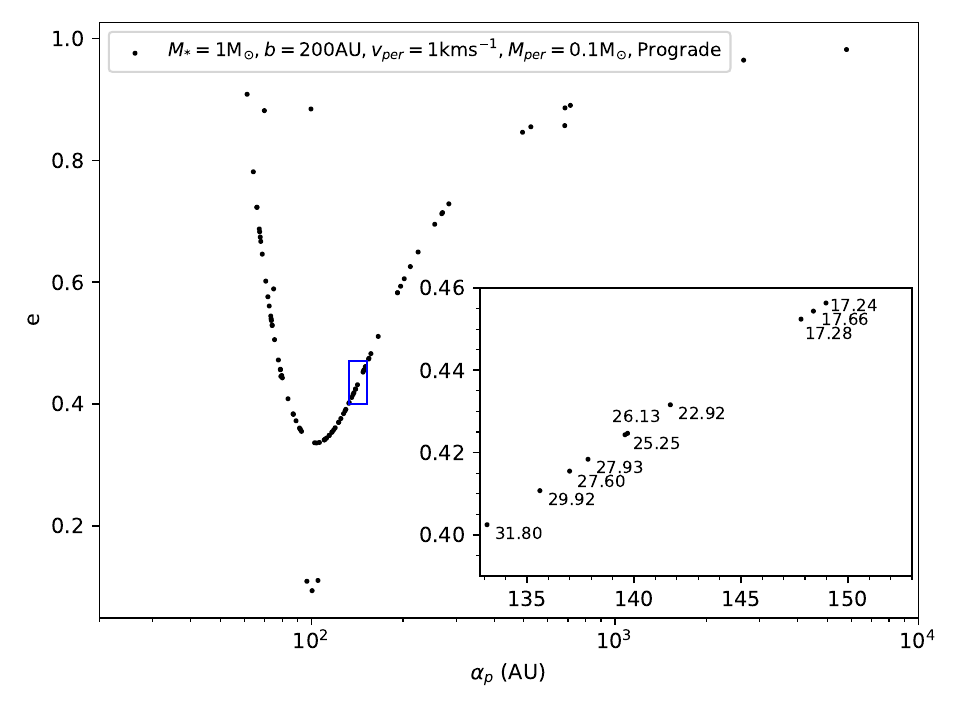}
    \caption{{The final eccentricity against the final semi-major axis of the bound population post-encounter for a specific set of initial parameters (as marked on the top graph). The subplot zooms in the section indicated by the blue rectangle, with the labels next to each point indicating the initial} true anomaly, $f$, of that planet (in degrees).}
    \label{fig:trueanomaly}
\end{figure}

We see distinct curves on the $a_p-e$ graphs (Figs.~\ref{fig:ecc_sm_b}-\ref{fig:ecc_sm_vel}) that are due to the fact that small variations of the initial true anomaly $f$ of the planet (we simulate 100 randomly chosen true anomalies per each combination of parameters) lead to small variations in the final configuration of the planetary system. This is demonstrated in Fig.~\ref{fig:trueanomaly}, where we select a small region of the  $a_p-e$ graph (final values) for a specific set of parameters (as seen on the top of the graph) and mark the initial true anomaly. We see that neighbouring points have similar initial true anomalies. Such patterns are not seen in simulations of planetary systems in clusters \cite [e.g.][] {Parker:2012, Zheng:2015g} as more bodies are involved in those, making them more chaotic (we simulate the interaction of only 3 bodies).

\bsp	
\label{lastpage}
\end{document}